\documentclass{jetpl}
\twocolumn
\lat

\title{Monte Carlo study of first-order transition in  
Heisenberg fcc antiferromagnet}
\rtitle{First-order transition in 
Heisenberg fcc antiferromagnet}
\sodtitle{Monte Carlo study of first-order transition in 
Heisenberg fcc antiferromagnet}

\author{M.\,V.\,Gvozdikova$^+$ and 
M.\,E.\,Zhitomirsky$^*$\/\thanks{mike.zhitomirsky@cea.fr}}
\rauthor{M.\,V.\,Gvozdikova, M.\,E.\,Zhitomirsky}
\sodauthor{M.\,V.\,Gvozdikova, M.\,E.\,Zhitomirsky}

\address{
$^+$Grenoble High Magnet Field Laboratory, CNRS, 38042 Grenoble, France \\
and Department of Physics, Kharkov National University, 61077 
Kharkov, Ukraine \\[1mm] 
$^*$Commissariat \`a l'Energie Atomique, DSM/DRFMC/SPSMS, 38054 Grenoble,
France}
\dates{\today}{*}

\abstract{Nearest-neighbor Heisenberg antiferromagnet on a face-centered 
cubic lattice is studied by extensive Monte Carlo simulations in zero 
magnetic field. The parallel tempering algorithm is utilized, which allows 
to overcome a slow relaxation of the magnetic order parameter and fully 
equilibrate moderate size clusters with up to $N\simeq 7\times 10^3$
spins. By collecting energy and order parameter histograms on clusters
with up to $N\simeq 2\times 10^4$ sites  
we accurately locate the first-order transition point at 
$T_c=0.4459(1)J$.}
\PACS{75.10.Jm,   
      75.50.Ee    
}

\begin{document}

\maketitle

\begin{figure}
\centerline{
\includegraphics[width=0.6\columnwidth]{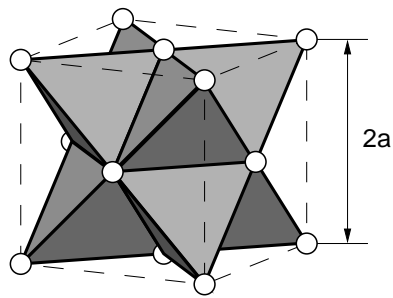}}
\caption{Fig.1: Face-centered cubic lattice as a network of edge-sharing
tetrahedra
}
\label{fcc}
\end{figure}

Geometric frustration generally denotes inability of a magnetic 
system to find unique classical ground state. It arises due 
to a competition between exchange interactions for certain 
types of magnetic sublattices. The wellknown examples are triangular 
and kagom\'e lattices in two dimensions and pyrochlore and 
face-centered cubic (fcc) lattices in three dimensions. 
Intriguing behavior of geometrically frustrated magnetic 
materials have attracted a lot of experimental and theoretical 
attention in the past decade \cite{hfm}.
Frustrated properties of Ising antiferromagnet on an fcc lattice 
(fig.\ref{fcc}) have been recognized a long-time ago \cite{ising}.
The case of vector (Heisenberg) spins has been investigated
to a lesser extent.
Experimental realizations of fcc magnets include type-I 
antiferromagnet UO$_2$ \cite{frazer,faber} and type-II antiferromagnet
MnO \cite{mno}.

In the present work we investigate a nearest-neighbor Heisenberg
antiferromagnet on an fcc lattice described by the Hamiltonian
\begin{equation}
{\cal H} = J \sum_{\langle ij\rangle} {\bf S}_i\cdot {\bf S}_j \ ,
\label{H}
\end{equation}
where ${\bf S}_i$ is a classical vector spin of unit length.
Every spin interacts with 12 nearest neighbors separated
by $(\pm a,\pm a,0)$, $(0,\pm a,\pm a)$ and $(\pm a,0,\pm a)$.
Frustrated properties of the model (\ref{H}) become apparent
if one calculates the mean-field transition temperature 
$T^{\rm MF}_c =\frac{1}{3}
|\min\{J_{\bf q}\}|$. The Fourier transform of exchange interaction
is
\begin{eqnarray}
J_{\bf q} &=& 4J\bigl[ \cos(q_xa)\cos (q_ya) + \cos (q_ya)\cos (q_za) 
\nonumber \\
& & \mbox{} \ \ \  +\cos (q_za)\cos (q_xa)\bigr] \ .
\label{Jq}
\end{eqnarray}
The minimum is reached at ${\bf Q} = (\pi/a)(1,q,0)$ with arbitrary $q$
and all equivalent 
wave-vectors in the cubic Brillouin zone. This set includes, in particular,
the type-I antiferromagnetic structure with
${\bf Q}_1=(\pi/a)(1,0,0)$ and 
the type-III
ordering with ${\bf Q}_3=(\pi/a)(1,\frac{1}{2},0)$.
The type-I (type-III) structure becomes the only
absolute minimum of $J_{\bf q}$ if a weak second-neighbor exchange
is added to Eq.~(\ref{H}) with ferromagnetic 
(antiferromagnetic) sign, see, {\it e.g.}, \cite{henley}.

Degeneracy of the nearest-neighbor model (\ref{H}) is a consequence
of decomposition of an fcc lattice into a network of edge-sharing
tetrahedra, such that every site is shared between 8 tetrahedra.
The classical ground state constraint consists, 
then, in a requirement of 
zero total spin for every tetrahedron.
This yields an infinite number of collinear and noncollinear
state with different periodicity, all of them having the same classical
energy. The harmonic spin-wave analysis shows that at low 
temperatures thermal fluctuations select collinear states
by an order by disorder effect \cite{henley}.
Such a result can be most easily understood by a method suggested
in Ref.~\cite{canals}, where the effect of 
short wave-length thermal fluctuations is shown to lead to 
an effective biquadratic exchange between neighboring spins. 
For the classical spins on an fcc
lattice thermal fluctuations generate the 
following low-temperature correction to the free-energy:
\begin{equation}
\Delta F = - \frac{T}{32} \sum_{\langle ij\rangle}
({\bf S}_i\cdot {\bf S}_j)^2 \ .
\end{equation}
Such a biquadratic interaction favors collinear spin 
arrangement. 
Two examples of collinear ground states include type-I spin
structure ${\bf S}_i = \hat{\bf e}\cos({\bf Q}_1\cdot{\bf r}_i)$ 
and type-III configuration 
${\bf S}_i = \hat{\bf e} \cos({\bf Q}_3\cdot {\bf r}_i+\frac{1}{4}\pi)$. 
Additional collinear ground states are constructed
from the above two configurations by selecting 
crystal planes (parallel to one of the cubic axes) with 
the N\'eel type of spin order and rotating all spins in such
planes by 180$^\circ$. All obtained collinear states have exactly
the same free-energy in the harmonic approximation
due to an extra gauge symmetry of a quadratic Hamiltonian
\cite{canals}. Their degeneracy is lifted by anharmonicities
in the spin-wave Hamiltonian---the problem, which to our knowledge
has not been considered analytically.

Finite temperature transition of a type-I fcc antiferromagnet
has been studied by the renormalization group
approach \cite{brazovskii,mukamel}.  Absence of stable
fixed points within the $\epsilon$-expansion suggests a 
first-order transition driven by thermal fluctuations.
The above calculations have been performed for the case
when the type-I structure corresponds to the absolute minimum,
that is the case of the spin model (\ref{H}) with a significant
ferromagnetic second-neighbor exchange.
In the nearest-neighbor case the anomalous contribution of
thermal fluctuations is further enhanced due to extra
soft modes.
Therefore, the conclusion about a first-order transition
should remain essentially unchanged.

Numerical Monte Carlo (MC) simulations of the finite temperature phase
transition in a nearest neighbor fcc antiferromagnet
have been performed by a standard single spin-flip technique
\cite{minor,diep89,alonco}.
First-order nature of the transition at $T_c\simeq 0.45J$
was unambiguously established
from the finite size scaling of the peak in the specific heat
\cite{diep89} and has been further supported by 
the energy histograms collected at the transition point \cite{alonco}.
As for the type of magnetic ordering at low temperatures there is no 
consensus among different authors. The first works have suggested 
the collinear type-I antiferromagnetic structure \cite{minor,diep89}, 
though no results for the magnetic structure factor have been produced.
In the subsequent more detailed study \cite{alonco} the low-temperature phase
of an fcc antiferromagnet was described as `a collinear state with
glassy behavior.' Apparent difficulty in simulation 
of an fcc antiferromagnet at low temperatures comes from the above
mentioned degeneracy between various collinear states,
which correspond to different local minima of the free-energy functional.
The collinear states
transform into each other under rotation of all spins in one
crystal plane. The local minima of the free-energy are, therefore,
separated by rather large entropy barriers $\sim L^2$, $L$ being a linear
size of the system. 
A single spin-flip MC technique cannot produce
appreciable moves in the phase space between distinct collinear 
configurations.
As a result, the spin system hardly relaxes to the absolute minimum and
the magnetic structure factor
does not exhibit good averaging 
in a reasonable
computer simulation time.

In the present work we shall apply the novel exchange MC method \cite{hukushima}
for simulation of a Heisenberg fcc antiferromagnet. This modification
of the standard MC technique, also called parallel tempering \cite{marinari},
has been developed for spin glasses, which is an outstanding 
example of hardly relaxing spin systems
with numerous local minima separated by macroscopic energy barriers.
In the exchange MC technique, several replicas
of the spin system are simulated in parallel
at a preselected set of temperatures. After a few
ordinary MC steps 
performed on each replica, 
adjacent in  temperature replicas are exchanged 
with a specially chosen probability \cite{hukushima}. 
The ensemble of parallel tempering, thus, includes
two Markov processes: stochastic motion in the multidimensional
phase space
of the spin model and a random walk along
a one-dimensional array of replicas. The second auxiliary 
process helps to dramatically decrease the correlation 
time for the main stochastic motion by repeatedly heating 
a given replica to high temperatures, where it quickly 
looses memory about the low-temperature magnetic structure.
The exchange MC technique allows to equilibrate moderate size 
systems and has gained popularity in simulation of spin 
glasses \cite{katzgraber} and frustrated Ising models \cite{bernardi}.
This method has not been applied so far to geometrically frustrated
vector spin models.

We have performed exchange MC simulations of the model (\ref{H})
on finite clusters with periodic boundary conditions
and $N=4L^3$ sites for $L=4$, 6, 8, 10, and 12
($N=256$--6912).
The highest temperature for the exchange MC ensemble
was selected at $T_{\rm up} = 0.75J>T_c$, where
the system equilibrates fast with a standard MC technique.
The lowest temperature in our simulations was $0.1J$.
The intermediate temperatures have been determined empirically
starting with a geometric progression in such a way
that the exchange rates for replicas swaps 
are roughly uniform and exceed 0.1 for the largest clusters.
In particular, in the vicinity of the first-order transition
point the temperature steps have to be 
decreased in order to avoid a bottleneck for replicas
drifts across $T_c$.
In total we have employed $N_T=72$ replicas for most cluster sizes.
The MC procedure for simulation of individual replicas
is the multiflipping Metropolis method, see, {\it e.g.}, \cite{diep89}, 
which includes several flipping attempts ($m=5$ in our case)
on every spin under the same local field
before moving to a next spin. This method is especially suitable
for an fcc lattice with a large coordination number $z=12$,
when calculation of local fields takes a significant part
of the CPU time. After sweeping 2 times over the whole cluster,
the replica exchanges have been performed. 
The above procedure constitutes one exchange MC step.
All replicas  were set in random initial configurations
and equilibrated  over $10^5$ exchange MC steps.
Measurements have been performed for additional 
$5\times 10^5$--$4\times 10^6$ exchange MC steps.
Equilibration of various physical characteristics has been checked 
(i) by absence of time evolution over at least a half of measuring
time, and (ii) by comparison to the results for a different set 
of initial replica configurations obtained by gradual annealing 
of a single replica from high temperatures.

\begin{figure}
\centerline{
\includegraphics[width=0.8\columnwidth,angle=0]{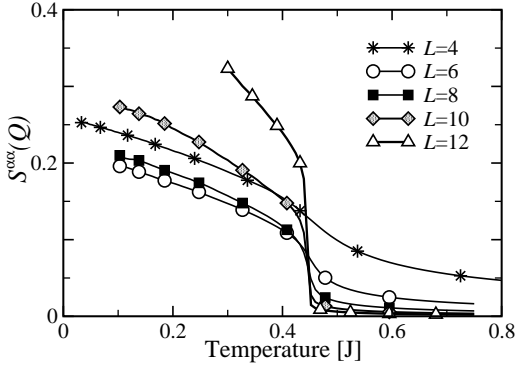}}
\caption{Fig.2: Temperature dependence of the structure
factor corresponding to the type-I antiferromagnetic ordering 
}
\label{op}
\end{figure}

During simulations of the exchange MC ensemble
the internal energy, the specific heat,
the magnetic structure factor, and the collinear order parameter
have been measured. Results for the internal energy and the specific
heat are identical with the previous results obtained
by the standard MC method \cite{diep89,alonco}
and are not discussed here. 
The magnetic structure factor is given by a square of the 
antiferromagnetic order parameter:
\begin{equation}
S^{\alpha\alpha}({\bf q}) = \frac{1}{N^2} \sum_{i,j} 
\langle S^\alpha_i S^\alpha_j \rangle e^{i{\bf q}({\bf r}_i-{\bf r}_j)} \ , 
\label{Sq}
\end{equation}
where $\langle\ldots\rangle$ denotes a thermal average.
For ${\bf q}={\bf Q}_3$ the structure factor exhibits a rather weak
$T$-dependence and scales as $1/N$ with increasing cluster
size at all temperatures. This generally indicates an 
absence of the corresponding order in 
the thermodynamic limit at any $T$.

Results for the structure factor of the type-I antiferromagnetic
ordering are presented in fig.\ref{op}. Contributions from 
three nonequivalent wave-vectors of the type-I structure  are added
together.
Symbols are used to distinguish different curves, while the lines
are drawn through the actual data for $N_T=72$ replicas.
For the largest cluster with $N=6912$ spins only $N_T=39$ 
replicas down to $T=0.3J$ have been equilibrated.
The remarkable feature of the presented data is an {\it inverse}
finite-size scaling at temperatures below $T_c\simeq 0.45J$: the
order parameter increases with the system size. 
The equilibrium sublattice magnetization deduced from 
$S^{\alpha\alpha}({\bf q})$ is still significantly smaller than 1 even
at $T=0.1J$. This is a consequence of thermally excited stacking faults,
domain walls and other defects in an ideal type-I antiferromagnetic structure.
Linear size of such defects coincides with the lattice size. 
Their concentration, therefore, decreases with increasing system
producing a significant increase of the order parameter.
Unfortunately, the lattice sizes are still too small to observe an asymptotic
thermodynamic behavior for $S^{\alpha\alpha}({\bf q})$, though they definitely point at
the spin ordering with the wave-vector of the type-I antiferromagnetic
structure.
The data for two large clusters exhibit a clear jump at the transition
temperature indicating the first-order transition.

\begin{figure}
\centerline{
\includegraphics[width=0.8\columnwidth,angle=0]{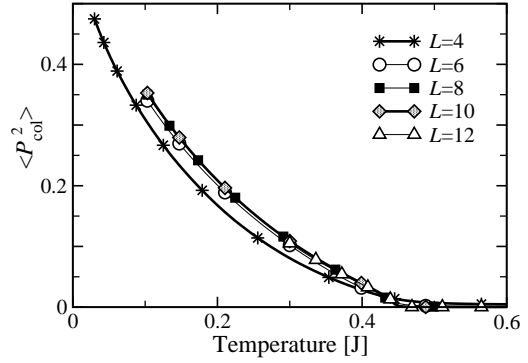}}
\caption{Fig.3: Temperature dependence of the collinear (nematic) 
order parameter
}
\label{nematic}
\end{figure}

Measurements of $S^{\alpha\alpha}({\bf q})$ cannot distinguish between 
presence of three domains of a single-$q$ type-I structure, which is always
the case for finite clusters, and  a noncollinear triple-$q$ 
spin state.
In order to study this aspect of the magnetic ordering in
an fcc antiferromagnet we define a collinear order parameter.
This is a single-site characteristics, which is insensitive
to periodicity (wave-vector) of the magnetic structure but
describes instead breaking of a spin-rotational symmetry.
The collinear order parameter is given by a traceless second-rank tensor:
\begin{equation}
P^{\alpha\beta}=\frac{1}{N}\sum_i \langle S_i^\alpha S_i^\beta\rangle 
- \frac{1}{3}\,\delta^{\alpha\beta}\ . 
\label{coll}
\end{equation}
It vanishes for a noncollinear triple-$q$ structure and has nonzero
value for a collinear state.
In the present case $P^{\alpha\beta}$ is a secondary order parameter: 
it couples linearly to a square 
of the primary antiferromagnetic parameter. 
For an $XY$ antiferromagnet
on a checkerboard lattice,  
$P^{\alpha\beta}$ is, however, a primary order parameter 
and characterizes a spin-nematic state \cite{canals}.

In MC simulations, a square of the order
parameter (\ref{coll}) is measured, which is given by
\begin{equation}
P_{\rm col}^2 = \frac{1}{N^2} \sum_{i,j}\langle 
S_i^\alpha S_i^\beta S_j^\alpha S_j^\beta\rangle 
- \frac{1}{3}\ . 
\label{coll2}
\end{equation}
The corresponding results are shown in fig.\ref{nematic}.
Since $P_{\rm col}^2$ is proportional to the fourth power
of an antiferromagnetic order parameter, it does not
show an appreciable jump at the first-order transition temperature. 
The data for the collinear order
parameter exhibit a remarkable lack of finite size scaling.
This indicates that spins over the whole lattice are predominantly
parallel or antiparallel to a certain direction. At $T=0.1J$
the aligned component of spin is $\langle S^\alpha\rangle \approx 0.78$.
Thus, the combination of the structure factor results (fig.\ref{op})
and the collinear order parameter data (fig.\ref{nematic}) points uniquely
to the collinear type-I antiferromagnetic order in a Heisenberg fcc
antiferromagnet.

\begin{figure}
\centerline{
\includegraphics[width=0.8\columnwidth,angle=0]{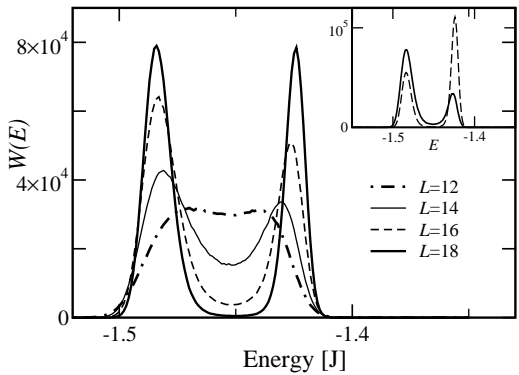}}
\caption{Fig.4: Distribution function (energy histogram) for several lattice 
sizes at the
first order transition $T_c=0.4459J$. Inset shows energy histograms for the
largest cluster $L=18$ ($N=23328$) at $T=0.4458J$, full line, and
$T=0.446J$, dashed line.
}
\label{histE}
\end{figure}

Finally, we present results for the energy and the order parameter 
histograms (distribution functions) at the transition point.
Since an fcc antiferromagnet exhibits a weak first-order transition,
we find it more advantageous to perform histogram collection with
a single replica instead of setting up an exchange MC ensemble.
Histograms have been collected utilizing a hybrid MC algorithm: 
3 multiflipping Metropolis steps are combined with 11 overrelaxation 
(microcanonical) updates \cite{overrelax}. On average 
$2\times 10^6$ configurations have been generated to build one 
histogram. Its accuracy has been checked by comparing the final 
distribution function to intermediate distributions obtained with 
proper rescaling from $10^6$ and $5\times 10^5$ configurations. 
The quality of the data shown in fig.\ref{histE} is significantly
higher than in the previous work \cite{alonco}, which allows us to
locate more precisely the transition point.
The first-order nature of the transition is clearly observed from
a double-peak structure of the energy histogram, see fig.\ref{histE}.
Positions of the two peaks do not change significantly with growing 
cluster size, while probability for intermediate states rapidly
drops with increasing $L$. At $T_c\approx 0.4459J$, the probability weights 
in the two peaks for the largest $L=18$ cluster ($N=23328$) differ 
by approximately $15\%$. To demonstrate a strong temperature dependence
of the relative weight of two peaks
we present on the inset of fig.\ref{histE} the distribution
functions at $T=T_c\pm 0.0001J$. The system clearly spends more time in the
upper (lower) peak above (below) the transition temperature.
We conclude, therefore, that the first-order transition
in a Heisenberg fcc antiferromagnet takes place at $T_c=0.4459\pm 0.0001J$.

\begin{figure}
\centerline{
\includegraphics[width=0.8\columnwidth,angle=0]{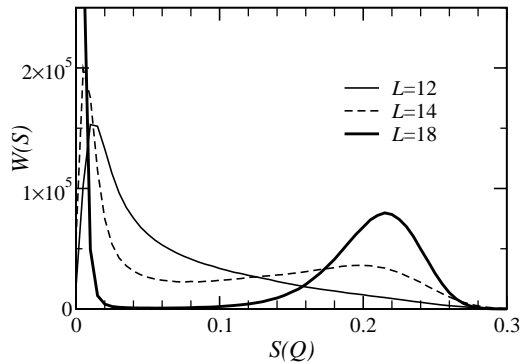}}
\caption{Fig.5: Histogram for the magnetic structure factor
at the
first order transition $T_c=0.4459J$
}
\label{histS}
\end{figure}

The distribution function for a square of the antiferromagnetic
order parameter is shown in fig.\ref{histS}. 
The largest cluster also develops a double-peak structure characteristic
for the first-order transition. Smaller clusters have, however,
significantly wider distributions for the magnetic structure factor,
which is not surprising in view of many local minima for different
collinear states.  It is interesting to notice that the jump
of $S^{\alpha\alpha}(Q)$ at the transition point determined from the 
histogram for the $L=18$ cluster (fig.\ref{histS})
is close to the jump observed for the $L=12$ cluster
in temperature scan (fig.\ref{op}), though the latter
lattice does not exhibit any visible double-peak structure at $T_c$. 
This indicates that apart from a close vicinity of the transition point 
the $L=12$ cluster may be very close to the thermodynamic limit.

In conclusion, we have performed Monte Carlo simulations for 
a finite temperature transition in a nearest-neighbor Heisenberg fcc antiferromagnet.
The obtained results clearly demonstrate a first-order transition
into a collinear type-I antiferromagnetic structure
due to an `order by disorder' effect.
Entropy mechanism for selection of the magnetic ordering
suggests an interesting sequence of phase transitions
for a weak antiferromagnetic second-neighbor exchange
$0<J'\ll T_c$. The higher temperature transition from
a paramagnetic state is determined by thermal fluctuations and
takes place into the type-I collinear
antiferromagnetic structure. At low enough temperatures $T\sim J'$
the energy selection overcomes the entropy effect
and a second transition from the type-I
into the type-III structure occurs.

We are grateful to H.\,G.\,Katzgraber for helpful comments
on the exchange MC method.
MEZ acknowledges hospitality of
the Condensed Matter Theory Institute of Brookhaven
National Laboratory during the course
of this work.

\end{document}